\documentclass{iopjournal}
\usepackage{graphicx} 
\usepackage{amsmath}
\usepackage{verbatim}
\usepackage{gensymb}
\usepackage{setspace}
\usepackage[noabbrev]{cleveref}
\usepackage{subcaption}
\usepackage{svg}

\usepackage{peer-review}
\newclipboard{line-numbers}

\begin{document}
\articletype{Paper}

\title{Homophily and wealth inequality shape mitigation behavior in coupled social-climate models}

\author{Luke Wisniewski$^1$\orcid{0009-0008-5816-1620}, Thomas Zdyrski$^{1,*}$\orcid{0000-0003-3039-172X} and Feng Fu$^{1,2}$\orcid{0000-0001-8252-1990}}

\affil{$^1$Department of Mathematics, Dartmouth College, Hanover, NH 03755, USA\\
$^2$Department of Biomedical Data Science, Geisel School of Medicine at Dartmouth, Lebanon, NH 03756, USA}

\affil{$^*$Author to whom any correspondence should be addressed.}

\email{thomas.zdyrski@dartmouth.edu}

\keywords{Social-Climate Dynamics, Evolutionary Game Theory, Climate Change, Wealth Inequality}

\begin{abstract}
Understanding the role of human behavior in shaping environmental outcomes is crucial for addressing global challenges such as climate change.  Environmental systems are influenced not only by natural factors like temperature, but also by human decisions regarding mitigation efforts, which are often based on forecasts or predictions about future environmental conditions.  Over time, different outcomes can emerge, including scenarios where the environment deteriorates despite efforts to mitigate, or where successful mitigation leads to environmental resilience.  Additionally, fluctuations in the level of human participation in mitigation can occur, reflecting shifts in collective behavior.  In this study, we consider a variety of human mitigation decisions, in addition to the feedback loop that is created by changes in human behavior because of environmental changes.  While these outcomes are based on simplified models, they offer important insights into the dynamics of human decision-making and the factors that influence effective action in the context of environmental sustainability. This study aims to examine key social dynamics influencing society's response to a worsening climate. While others conclude that homophily prompts greater warming unconditionally, this model finds that homophily can prevent catastrophic effects given a poor initial environmental state. Assuming that poor countries have the resources to do so, a consensus in that class group to defect from the strategy of the rich group (who are generally incentivized to continue ``business as usual'') can frequently prevent the vegetation proportion from converging to 0.
\end{abstract}

\section{Introduction}
\begin{LineLabel}{ED-1}
Understanding how human behavior impacts climate change has been a subject of inquiry for an extended period of time, with implications in both physical \cite{manabe1975effects} and theoretical \cite{hasselmann1976stochastic} climate modeling. In recent years, however, focus has shifted to human behavior responsible for climate change, with focuses on understanding the forces behind anthropogenic climate change \cite{moore2022determinants}, steps towards successful mitigation practices \cite{bury2019charting}, as well as an understanding of the social dynamics underpinning climate skepticism \cite{bago2023reasoning}. Such research has led to further investigation into the relationship between human behavioral dynamics and environmental responses.
\end{LineLabel}

\begin{LineLabel}{test}
Many pre-existing models have been built in order to parse out important relationships such as those between vegetation and mitigative action \cite{fushu:23}, temperature and income inequality \cite{menard2021}, perceived environmental risk and mitigative action \cite{liu23}, and others. 
\end{LineLabel}
Given the plethora of parameters and assumptions baked into each of these models, applicability to the real world climate system is always considered as one of the primary benchmarks. Creating a simplified model which synthesizes many of the important components of previous theoretical models is hence seen as an optimal way to minimize potential errors in parameterization and instead draw broad generalizations from many initial states. 

Investigations of climate change solutions have taken many forms. Many studies have focused in developing climate projections from real-world climate data from over the last 20,000 years
\begin{LineLabel}{ED-2}
\cite{braconnot2012evaluation}. This paleoclimatic data has been effectively utilized to investigate the effect of warming on arctic climates \cite{anderson06}, humanity's effect on the climate \cite{marlon09}, the impact of climate change on the Global South \cite{bassinot11}, as well as anthropogenic changes to arctic climate \cite{kaplan03}.  
\end{LineLabel}
These lines of inquiry have led to the creation of incredibly complex and increasingly robust Earth-systems models \cite{lucas-picherargüeso2021}. In terms of investigating possible climate solutions, evolutionary game theory is commonly used to model important social dynamics
\begin{LineLabel}{ED-3}
to overcome tragedies of the commons \cite{chen2018punishment} or collective risk dilemmas \cite{jiang2023deterrence} which are of interest in analyzing the issue of climate change.
\end{LineLabel}
 Many have treated climate action as a social-cooperation game to avoid catastrophic results
 \begin{LineLabel}{ED-4}
, with analysis ranging from altering internal perceptions of climate risk \cite{barfuss2020caring}, interpersonal communication (such as measurements of future outlook \cite{wang2020communicating} and of cooperation \cite{Xiao_2023}), risk sharing \cite{santos2021dynamics} and the importance of understanding previously unforeseen risk \cite{kruczkiewiczklopp2021}, as well as exploring the role of central governance on climate issues \cite{levin2022governance} and the potential harms from long-term climate scenarios \cite{kempxu2022}.
 \end{LineLabel}
 Others have even prioritized experimental designs \cite{Alfonso_2024}. In many cases, these models involve interaction and feedbacks between human activity and the environmental state.
\begin{LineLabel}{ED-5}
    Implications have been far-reaching, including potential stability of oscillations between environmental states \cite{shufu2022}, delayed reward mechanisms for mitigators in social-climate games \cite{Szolnoki_2017}, the conditions required for environmental resiliency \cite{wang2020eco}, and the frequency of non-renewable resource depletion \cite{Muñoz-Álvarez_2024}.
\end{LineLabel}
While social-climate dynamics in general are a growing topic of study, the existent literature on inequality in these dynamics still is relatively scarce.
 \begin{LineLabel}{ED-6}
     \cite{RobiouDuPontEtAl2025} find that the largest global economies are most responsible for the gap between current and target CO2 emissions. Furthermore, \cite{Jiang_2025} discuss the effectiveness of cooperation incentives, demonstrating a negative correlation between cooperation and inequality. 
 \end{LineLabel}

Using a forest dieback model coupled to a homogeneous population, \cite{fushu:23} finds a limited number of stable fixed-points (varying depending on the initial temperature conditions), signifying a limited number of potential vegetative outcomes in equilibrium. This simplified model already limits the number of potential climate futures, but assumes homogenized incentives within the population, rather than allowing different individuals or groups to face different incentives to mitigate climate change. \cite{menard2021} accounts for population differences and dynamics. Modeling populations as divided into rich and poor groups (with interactions in between), they find through an Earth-systems model that increasing homophily (how much individuals want to only interact within their own group) increases peak temperature anomaly, regardless of initial environmental conditions. Given polarization on current climate issues, accounting for differing incentives is critical to understanding how to maximize engagement in mitigative efforts. 

In the study of social-climate models, interpretability is key
\begin{LineLabel}{ED-7}
    \cite{Trenary_DelSole_2022}, particularly as machine learning methods and models are increasingly used to predict climate outcomes such as precipitation \cite{Gibson2021} and ice melt \cite{ale2025advancingclimatemodelinterpretability}.
\end{LineLabel}
\begin{LineLabel}{R2-2}
    This paper contributes in tandem to previous work found in \cite{fushu:23} and \cite{menard2021}, but makes important novel contributions. First, coupling a heterogeneous population model \cite{menard2021} with a forest dieback model \cite{fushu:23} reduces concerns regarding the tuning of earth systems models, a persistent issue in earth systems modeling as contextualized by \cite{hourdinetal2017}. Through a simpler forest dieback model, issues regarding over- or mis-parameterization are avoided, while results maintain a high level of interpretability. Furthermore, this paper is interested in additional social factors contributing to climate responses, including costs of mitigation, effectiveness of mitigative action, perceived costs to climate change impacts, and social norm strength. Discussion of the effects of these social factors on long-term climate outcomes has so far not been addressed.
\begin{LineLabel}{R1-1.1}    
    Finally, this paper proposes a novel hypothesis: that present underlying social and environmental conditions are a key component of determining the best course of climate action. This paper hence provides an important, and testable, future direction of academic research.
\end{LineLabel}
\end{LineLabel}

 While a variety of Earth-systems models may more accurately represent the conditions governing our climate system, simplified models offer improved interpretability and enable parameter variation to highlight key relationships. Hence, this study primarily couples the two strong points of the methods explained above: an easily interpretable forest-dieback vegetative model with more robust social dynamics to account for pervasive societal inequalities. The rest of the paper will proceed as follows. Section 2 contains a discussion of our model design, section 3 describes first the main results of our simulations as well as additional robustness checks for our model. Section 4 then summarizes our findings and offers perspectives for further research.

\section{Methods and Model}

\subsection{Vegetation Model}

Discussions of forest dieback models frequently are centered around discussion of the Amazon rainforest, a critical marker for climate health~\cite{ritchie2021overshooting}. We represent the vegetation coverage, $v$, as a proportion between 0 (barren, no coverage) and 1 (lush, full coverage), where $v$ is governed by the following differential equation,
\begin{LineLabel}{R1-3}
    with $t$ measured in years
\end{LineLabel}
:
\begin{equation}
\frac{dv}{dt}=gv(1-v)-\gamma v
\label{eq:uncoupledveg}
\end{equation}
Here, $g$ is the growth rate, and $\gamma$ is the decay, or disturbance rate. As in \cite{ritchie2021overshooting}, $\gamma=0.2$. As growth of plants depends on how proliferated the plant already is, the growth rate $g$ is governed by the equation seen below:
\begin{equation}
    g=g_0\left[ 1-\left( \frac{T-T_{opt}}{\beta}\right)^2 \right]
    \label{eq:grate}
\end{equation}
Where $g_0=2$ is the maximum growth rate, $T_{opt}=28\degree$C is the optimal temperature for vegetation growth, and $\beta=10$ is the half-width of the growth versus temperature curve \cite{fushu:23}. Finally, $T$ represents the actual temperature, and is represented as follows:
\begin{equation}
    T = T_v + (1- v)a. 
    \label{eq:temp}
\end{equation}
Importantly, $T_v$ represents the temperature under full forest coverage conditions. 
\begin{LineLabel}{addingdetail}
As in \cite{fushu:23}, we call this term the ``ambient temperature''.
\end{LineLabel}
~Additionally, $a=5$ represents the difference between surface temperatures in barren and completely forested regimes---that is, the temperature cost of losing all vegetation in the forest.

\cite{fushu:23} contains a detailed explanation of this forest model in isolation, so we will postpone further discussion of the vegetation model to subsection 2.3, where the coupling with the social dynamic model will be detailed.

\subsection{Social Dynamics Model}
As in \cite{fushu:23}, \cite{menard2021}, and others concerning public goods games~\cite{wu2014social}, this study considers a population deciding between two possible actions: mitigation ($M$) and non-mitigation ($N$). As in \cite{menard2021}, this study for simplicity creates a population divided into two groups, which for convenience are called ``rich'' ($R$) and ``poor'' ($P$). These two groups differ in terms of the costs required to engage in mitigative action, the income they have to be able to engage in mitigative actions, their level of satisfaction with the climate, and size in the entire population. Additionally, between-group interaction is governed by a homophily term, representing the relative value which a player puts on the actions of those in the opposing group compared to those in the same group. This term will be notated as $h$. When $h=0$ (no homophily), players value the actions of the opposing group equally to those of the people in their own group. When $h=1$ (high homophily), players place no value on the actions of the opposing group and instead only place value on the actions of their own group.

Let the subscript $i\in\{0,1\}$ denote the class group, where $i=0$ indicates the rich group and $i=1$ indicates the poor group. Subscripts $R$ and $P$ will be used in later sections for clarity. For a given value of $i$, $j=1-i$ denotes the opposing group. For group $i$, $x_i$ represents the proportion of mitigators in that group while $y_i=1-x_i$ represents the proportion of non-mitigators. 
The payoffs for the group $i$ are
\begin{equation}
    E_i(M) = -\alpha_i + \frac{1}{2}f(T_f)+\delta[x_i+(1-h)x_j]
    \label{eq:payoffmitigator}
\end{equation}
for mitigators and
\begin{equation}
    E_i(N) = - \frac{1}{2}f(T_f)+\delta[y_i+(1-h)y_j]
    \label{eq:nonmitigator}
\end{equation}
for non-mitigators \cite{menard2021}. $\alpha_i$ represents the costs of mitigation for group $i$, and
\begin{LineLabel}{R1-5.1}
$\delta=1$ represents the strength of social norms, 
creating an incentive to align one's opinion with that of the group majority. The social-norm term $\delta$ acts by enhancing the mitigator payoff $E_i(M)$ by the mitigator proportion $\delta x_i$ and enhancing the non-mitigator payoff $E_i(N)$ by the non-mitigator proportion $\delta y_i$. This encourages polarization along the mitigator/non-mitigator axis, while the homophily term $h$ encourages polarization between poor/rich groups.
\end{LineLabel}
 ~Furthermore, $h$ is the homophily term described above, and $x_j$ represents the proportion of people in the opposing group who are mitigators. Finally, $f(T_f)$ is a sigmoid function representing the personal cost of projected global warming, and is represented by the following equation\footnote{See \cite{fushu:23} for a full derivation of this equation.}

\begin{equation}
    f(T_f)=\frac{f_{\max}}{1+e^{-w[-7.5v(t)+7.5v(t-10)-T_c]}}
    \label{eq:ftf}
\end{equation}
Here, $f_{\max}$ is the perceived maximum warming cost, $w$ is the perceived nonlinearity of warming cost, $v(t)$ is the vegetation proportion at time $t$ (more simply, the currently perceived vegetation), $v(t-10)$ is the perceived vegetation proportion ten years ago, and $T_c$ is the maximum temperature rise which can be tolerated by individuals, from here on called the critical temperature\footnote{We investigate the sensitivity of the results to the values of $f_{\max}$ and $w$ in section 3.4. When otherwise not specified, $f_{\max}=5$ and $w=3$.}. As explained by \cite{menard2021}, this cost is designed to be an explicit incentive to mitigate and a disincentive to not mitigate (a social benefit/cost), while the material costs of climate change would impact mitigators and non-mitigators equally within groups. 
\begin{LineLabel}{R1-5.2}
    Non-mitigators may experience psychological costs due to their inaction, while mitigators may experience positive personal externalities, such as personally improved health due to their changed behaviors. Hence, $f(T_f)$ is meant to capture the additional costs of inaction and benefits of climate action not captured through other channels.
\end{LineLabel}

\begin{LineLabel}{R1-5.3}
Furthermore, perceived unfairness may alter players' incentives. As explained in \cite{menard2021}, the importance of climate dissatisfaction relative to the influence of climate change is a key determinant of climate action. This inequality alters the costs of mitigation for the poor group, as seen below
\begin{equation}
    \alpha_P=\alpha_{P_0}+\frac{d}{1+e^{-\Omega\left(\frac{y_R}{y_{R0}}\frac{I_R}{I_P}\frac{I_{P0}}{I_{R0}}-d_c\right)}}
\end{equation}
Here, $\alpha_{P_0}$ represents the initial costs of mitigation for the poor group, $d$ is the maximum cost of dissatisfaction, creating larger costs for the poor group to mitigate as dissatisfaction with the rich group increases. $\Omega$ is the nonlinearity of dissatisfaction costs, while the ratio $\frac{y_R}{y_{R_0}}$ is the proportion of non-mitigators in the rich group normalized by the initial non-mitigation proportion. Finally, $I_R/I_P$ is the ratio of current incomes between rich and poor groups, normalized by the initial incomes\footnote{An explanation of the evolution of the wealth dynamics can be found in the appendix of \cite{menard2021}} $\left(I_{P_0}/I_{R_0}\right)$, and $d_c$ is the critical value of the dissatisfaction cost. As explained by \cite{menard2021}, dissatisfaction increases when there are more non-mitigators in the rich group, or as income inequality increases between the rich and poor groups. Thus, for the poor group, the relative importance between dissatisfaction versus the impact of climate change becomes a key factor in determining behavior. When dissatisfaction is low relative to the impact of climate change, incentives for mitigation are high, while when dissatisfaction is high relative to the impact of climate change, incentives for mitigation are low. 
\end{LineLabel}

Following the derivations of \cite{menard2021}, we obtain the following replicator dynamics

\begin{multline}
\frac{dx_i}{dt}= \kappa x_iy_i[E_i(M) - E_i(N)] + (1-h)\kappa\\\{x_j\max[E_j(M)-E_i(N), 0]y_i-y_j\max[E_j(N)-E_i(M), 0]x_i\}    
\label{eq:socialdynamicrate}
\end{multline}

Here, $\kappa=0.05$ represents the social learning rate, or how quick players are to change their opinions based on perceived information.

\subsection{Coupled Social-Climate Model}

While we divided the mitigation proportions into two groups, $x_R$ and $x_P$, from an environmental perspective the total mitigation effort, regardless of social divisions, is most important. To get the total mitigation proportion $x$, we define the following relation

\begin{equation}
    x=\min\{s\rho x_R+(1-\rho)x_P, 1\}
    \label{eq:xrate}
\end{equation}
Here $\rho=0.25$ represents the proportion of the total population which is in the rich group. Additionally, $s\geq1$ represents the relative effectiveness of mitigative action for the rich group. Not only might the rich group have lower costs to mitigating, but the opportunity for technological advancement implies that the rich group may be able to provide greater environmental benefit with less collective effort (corresponding to a larger $s$ value). 
\begin{LineLabel}{R2-v}
This equation is here newly defined to ensure proper coupling of the social and climate components of our model.
\end{LineLabel}

Following \cite{fushu:23}'s aim of adding a term $\eta(x)$ to the growth rate in the vegetative dynamics which increases monotonously as the total proportion of mitigators increases, this study sets $\eta(x)=0.2+0.4x$ and multiply to \cref{eq:grate} to get 
\begin{equation}
    g=g_0\left[1-\left(\frac{T-T_{opt}}{\beta}\right)^2\right]\eta(x)
    \label{eq:gratex}
\end{equation}
Plugging in the parameters as described above for all values except $T_v$, we get 
\begin{equation}
    \frac{dv}{dt}=2[1 - 0.01(T_v - 23 - 5v)^2](0.2 + 0.4x)v(1 - v) - 0.2v
    \label{eq:coupledveg}
\end{equation}
Hence, this study define the coupled delay differential equation system of three equations as follows\singlespacing
\begin{equation}
\begin{cases}
    \frac{dv}{dt}=2[1 - 0.01(T_v - 23 - 5v)^2](0.2 + 0.4x)v(1 - v) - 0.2v & (\text{i})\\
    \\
    \frac{dx_R}{dt}= \kappa x_Ry_R[E_R(M) - E_R(N)] + (1-h)\kappa\\\{x_P\max[E_P(M)-E_R(N), 0]y_R-y_P\max[E_P(N)-E_R(M), 0]x_R\} & (\text{ii})\\
    \\
    \frac{dx_P}{dt}= \kappa x_P y_P[E_P(M) - E_P(N)] + (1-h)\kappa\\\{x_R\max[E_R(M)-E_P(N), 0]y_P-y_R\max[E_R(N)-E_P(M), 0]x_P\}& (\text{iii})
\end{cases}
\label{eq:coupledmodel}
\end{equation}\doublespacing

A comprehensive list of parameters used is shown in \cref{tab:parameters} below. 
\begin{LineLabel}{R2-par}
When not explicitly referenced in the legend of following figures, the default values of parameters have been used.
\end{LineLabel}
\begin{table}[htb!]
    \centering
    \resizebox{\textwidth}{!}{
    \begin{tabular}{|c|c|c|c|}
    \hline
         \textbf{Variable Name} & \textbf{Variable Description} & \textbf{Default Value} & \textbf{Value Range}  \\\hline
         $x_R$ & Proportion of Mitigators in the Rich Group & & $[0,1]$\\
         $x_P$ & Proportion of Mitigators in the Poor Group & & $[0,1]$\\
         $v$ & Vegetation Proportion & & $[0,1]$\\
         $\rho$ & Proportion of the Rich Group in the Entire Population & 0.25 & $[0,1]$\\
         $h$ & Homophily Parameter & & $[0,1]$\\
         $\gamma$ & Vegetative Decay Rate & 0.2 &\\
         $g_0$ & Maximum Vegetative Growth Rate & 2 &\\
         $T_{opt}$ & Optimal Temperature for Vegetation Growth & 28 & \\
         $\beta$ & Half-Width of Growth Vs. Temperature Curve & 10 & \\
         $a$ & Temperature Cost of Losing All Vegetation & 5 & \\
         $f_{\max}$ & Maximum Warming Cost & 5 & $[1,10]$\\
         $w$ & Nonlinearity of Warming Costs & 3 & $[1,10]$\\
         $\alpha_{R_0}/\alpha_{P_0}$ & Ratio of Initial Mitigation Costs Between Rich and Poor Groups & 0.5 & $[0,1]$\\
         $\delta$ & Strength of Social Norms & 1 & $[1,4]$\\
         $d$ & Maximum Cost of Dissatisfaction & 5 & $[1,10]$\\
         $\kappa$ & Social Learning Rate & 0.05 & \\
         $s$ & Relative Effectiveness of Mitigative Action for Rich Group & 1 & $[1,10]$\\
         $I_{R_0}$ & Initial Income of Rich Group & 5 & \\
         $I_{P_0}$ & Initial Income of Poor Group & 3.5 & \\ 
         $T_c$ & Critical Temperature & & $[0,3]$ \\
         $T_v$ &Ambient Temperature & & $[10,35]$\\
         $\Omega$ & Nonlinearity of Dissatisfaction Cost & 3 &\\
         $d_c$ & Critical Value of Dissatisfaction Cost & 1.5 & -\\
         \hline
    \end{tabular}
    }
    \caption{List of Main Used Parameters}
    \label{tab:parameters}
\end{table}

\section{Results}
To complete the necessary simulations, we first perform a spin up period of 10 years with the vegetative level using the uncoupled model seen in \cref{eq:uncoupledveg}. Then for years 10-200, this study simulates using the delayed differential equation system of \cref{eq:coupledmodel}. Simulations were performed with the ddeint package in Python.

\subsection{Stability Analysis}
\begin{figure}[htb!]
    \centering
    \begin{subfigure}[t]{0.32\linewidth}
    \centering
        \includegraphics[width=1\linewidth]{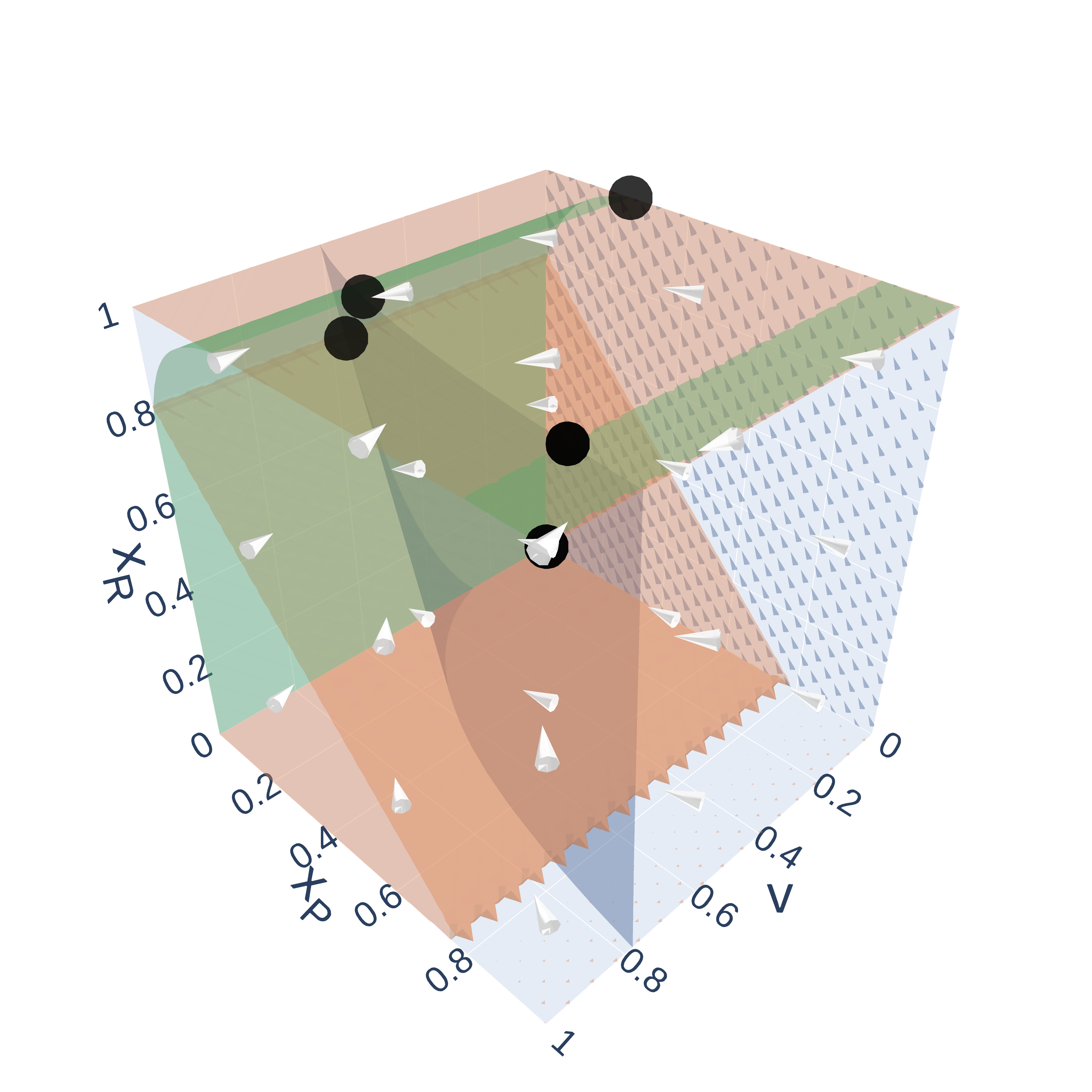}
    \caption{$h=0$}
    \label{fig:phaseportraith0}
    \end{subfigure}
    \begin{subfigure}[t]{0.32\linewidth}
    \centering
        \includegraphics[width=1\linewidth]{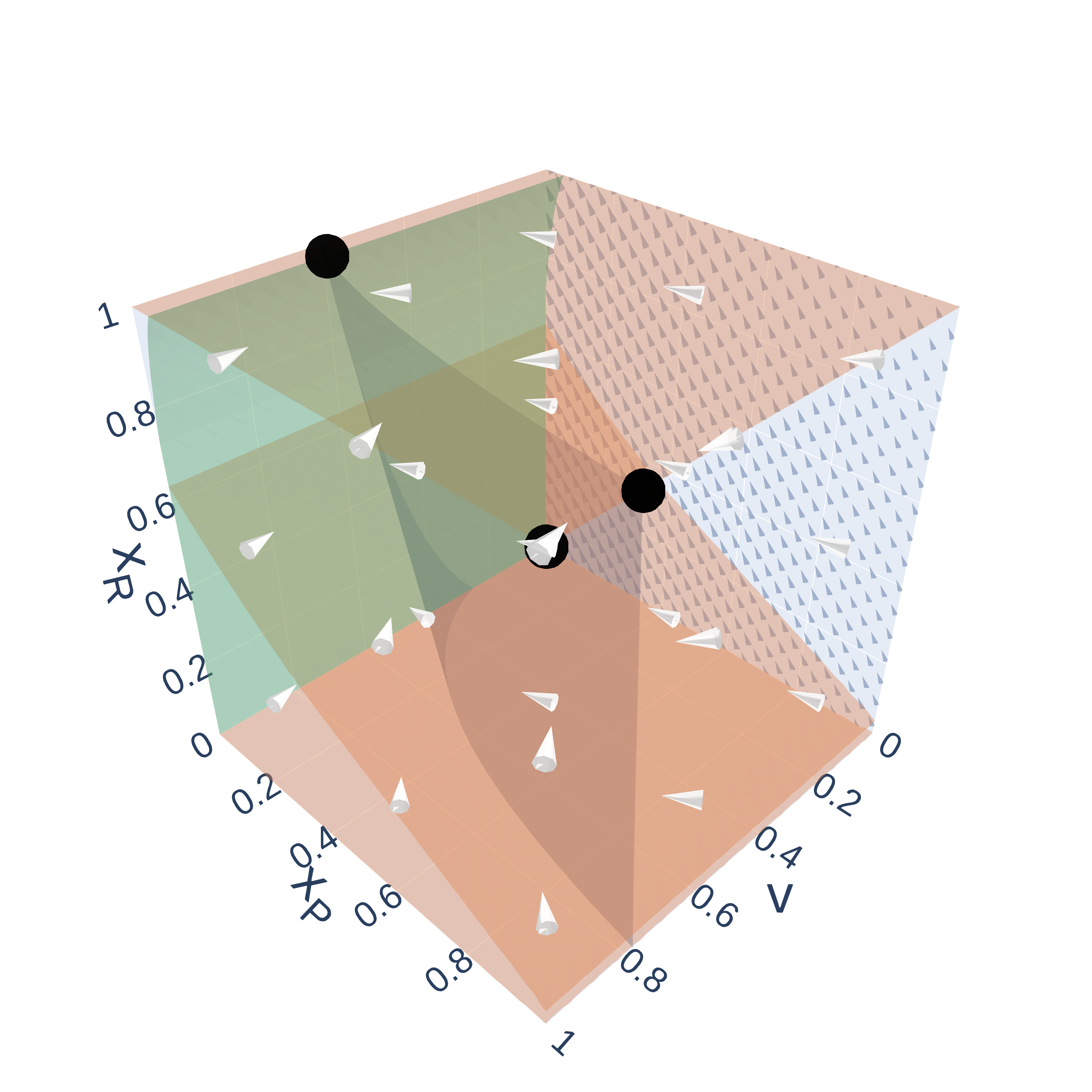}
    \caption{$h=0.5$}
    \label{fig:phaseportraith05}
    \end{subfigure}
    \begin{subfigure}[t]{0.32\linewidth}
    \centering
        \includegraphics[width=1\linewidth]{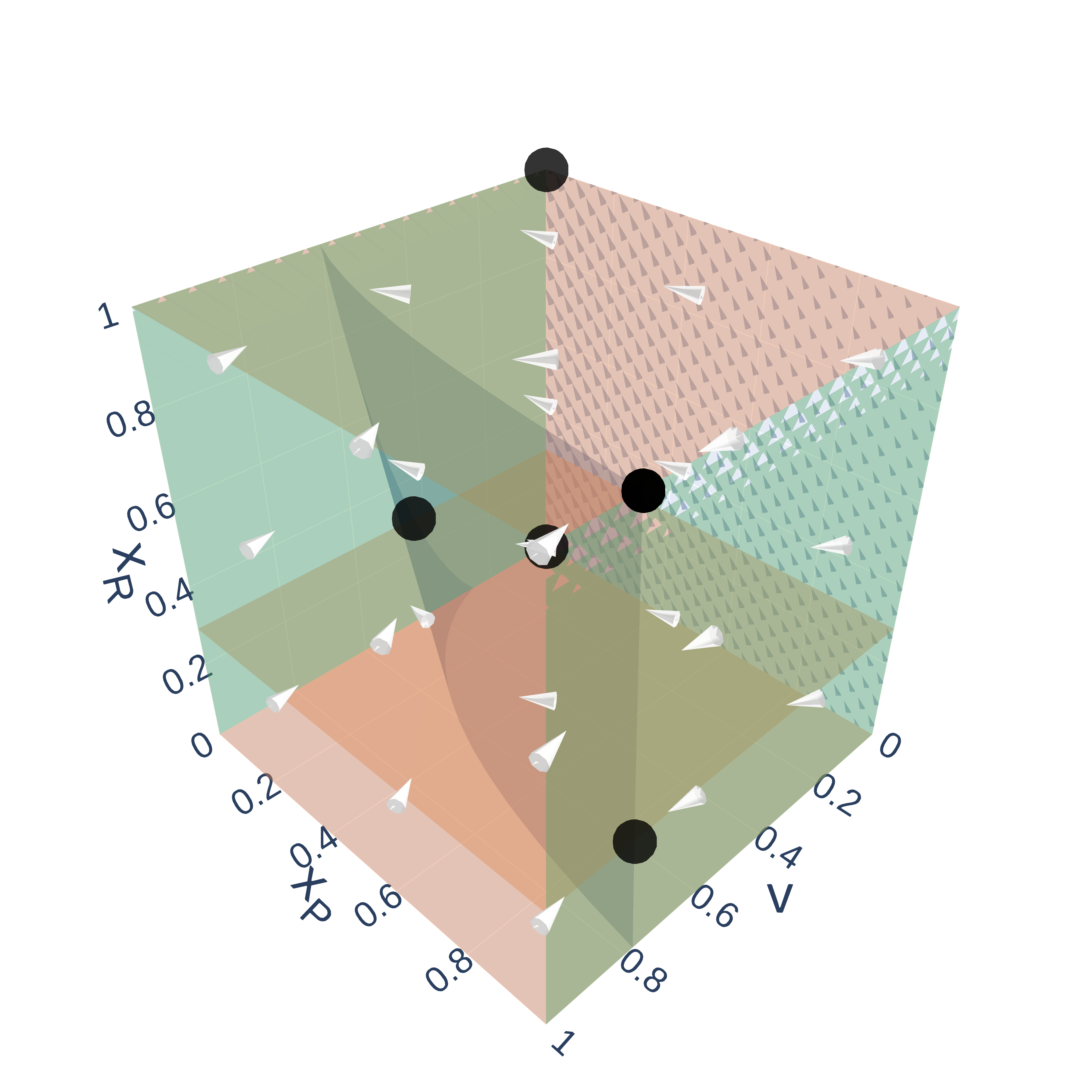}
    \caption{$h=1$}
    \label{fig:phaseportrait1}
    \end{subfigure}
    \caption{
        Phase Portraits over Varying Levels of Homophily.
        }
        \begin{minipage}{15cm}
        \footnotesize
        Notes: The above figures depict phase portraits of the system of equations over three different levels of homophily ($h$). \Cref{fig:phaseportraitfull} panel a depicts the system with no homophily, understood as no bias towards one's own groups. \Cref{fig:phaseportraitfull} panel b depicts the system under a moderate amount of homophily ($h=0.5$), while \cref{fig:phaseportraitfull} panel c depicts the system under full homophily ($h=1$). Black dots represent stable fixed points of the system of equations. The blue planes represent the $v$ nullclines, the green planes represent the $x_P$ nullclines, and the orange planes represent the $x_R$ nullclines. The white arrows in the plot represent directions of movement of the system. $T_v=30$ and $T_c=0.5$.
        \end{minipage}
    \label{fig:phaseportraitfull}
\end{figure}

To get a sense of how the system of equations depends on the homophily term, phase portraits were developed over varying levels of homophily, seen in \cref{fig:phaseportraitfull}.
\begin{LineLabel}{R1-7}
    To create these phase portraits we used the Python plotly library to graph the three isosurfaces, representing the planes where one of the three rate equations of \cref{eq:coupledmodel} is equal to zero. Then, we used Python's root solver fsolve to determine the fixed points of the system,
    represented by black dots. 
\end{LineLabel}

While \cite{fushu:23} finds a number of fixed points which vary by the temperature parameters, this study also finds that the stability of the system depends on the homophily level. Here the blue planes represent nullclines of the vegetation proportion $v$, green the proportion of poor mitigators $x_P$, and orange the proportion of rich mitigators in the population $x_R$. The equilibrium points to our system of equations exist at the intersections of these three planes, and the arrows seen within the plots indicate the direction of movement of the system to a given fixed point. Given the shape of the $x_P$ nullclines, one can begin to see an uneven weighting of opinions between the rich and poor groups. When $h=1$, there is a (vertical) nullcline at $x_P=1$, while at the other values shown, this sheet is only seen at a very high level of $x_R$. When $h<1$, the poor group requires a large mitigative effort from the rich group in order to converge to a high mitigation proportion themselves. On the international stage, many small nations opt to follow the directives of global powers; hence our model appears to capture a similar phenomenon. Due to the complexity of the system of equations, a simple closed-form solution to all three equations is not possible without significant parameterization. However, one can notice an identical shape for the $v$ nullclines as those seen in \cite{fushu:23} in all three phase portraits when looking on the $v-x_P$ plane.

\subsection{Case 1: Increasing Homophily Worsens Environmental Outcomes}

\cite{menard2021} find unambiguously that increasing the homophily parameter increases peak temperature anomaly in their earth systems model. In the coupled forest dieback model, we find analogous results given particular underlying conditions. \Cref{fig:heatmap_favorable_conditions} shows a heatmap, where the color scheme represents the vegetation proportion after simulating for 200 years for given values of $T_v$ (horizontal axis) and $T_c$ (vertical axis). The dark brown regions represent simulations where the vegetation has completely died out, while the light green color represents a larger vegetation proportion after the 200 year simulation period. When $h=0$, there exists a large range of $T_v$ and $T_c$ values which result in a very high vegetation 
\begin{LineLabel}{R2-ii}
proportion,  
\end{LineLabel}
around 0.8. However, when $h=1$, this region becomes very small, and most $T_v$ and $T_c$ values result in a vegetation proportion around 0.5.

Interestingly, \cref{fig:heatmap_favorable_conditions} panel a closely mirrors figures seen in \cite{fushu:23}. This has two main implications. Firstly, the deterministic models seen here and in \cite{fushu:23} allow for very few unanticipated outcomes. \cite{fushu:23}, considering both the vegetation proportion and the mitigation proportion, occasionally (but rarely) find outcomes that seem unintuitive (where the forest survives but no one is a mitigator) or the nature of the system (where convergence is not achieved after 200 years). The system here also does not support oscillatory equilibria, so the presence of oscillations suggests some transients take longer than the 200 year simulation period to converge. However, such cases are not found in our results.

The other main implication stems from the nature of the system of equations itself. When $h=1$, dynamics very close to the more simplistic dynamics seen in \cite{fushu:23} are obtained. However, it is when $h=0$ when the similar results can be seen. This implies that removing homophily from the system results in a convergence of opinions, resulting in a uniform population like the one analyzed in \cite{fushu:23}. While under this set of parameters an agreement to mitigate is reached among the population, we will later see that agreement is not always reached in favor of mitigation. 

\begin{figure}[htb!]
    \centering
    \includesvg[width=1\linewidth]{Figures/fmaxgoodcombined.svg}
    \caption{Final Forest Coverage Under Favorable Conditions For Mitigation}
     \begin{minipage}{15cm}
    \footnotesize
        Notes: The above figures depict heatmaps of forest coverage $v$ after 200 years under different ambient temperature $T_v$, critical temperature $T_c$, and homophily level $h$. \Cref{fig:heatmap_favorable_conditions} panel a depicts the long term vegetative outcomes when $h=0$, while \cref{fig:heatmap_favorable_conditions} panel b depicts the long-term outcomes when $h=1$. To define a regime where mitigation is favored, the perceived cost of climate change is high ($f_{\max}=6$), while the cost of expressing dissatisfaction over the state of the climate is low ($d=1$). $v_0=0.1, x_{R_0}=0.1, x_{P_0}=0.8$.
    \end{minipage}
    \label{fig:heatmap_favorable_conditions}
\end{figure}

\Cref{fig:favorablesliders} presents the changes over time of the main parameters of interest, $x$, $x_R$, $x_P$, and $v$, given particular values of $T_v$ and $T_c$. Here, one can clearly see that prohibiting inter-group discussion results in the group with less to lose from climate change (the rich group) lacking the appropriate social push to participate in mitigative activities. \Cref{fig:favorablesliders} panel a shows the scenario without homophily, while \cref{fig:favorablesliders} panel b displays the time trends under a population with total homophily. As one would expect, the convergence of opinions to agreeing upon maximum mitigation efforts ($h=0$) results in a better environmental state in the long-run.
\begin{figure}[htb!]
    \centering
    \includesvg[width=1\linewidth]{Figures/goodsliderscombined.svg}
    \caption{Trajectories Over Time Under Favorable Conditions For Mitigation}
    \begin{minipage}{15cm}
    \footnotesize
    Notes: The same initial vegetative state and mitigation proportions under different levels of homophily. \Cref{fig:favorablesliders} panel a shows the values of vegetation $v$, the mitigation proportions for the rich and poor groups, $x_R$ and $x_P$, respectively, as well as the total mitigation proportion, $x$, when $h=0$ over the course of the 200 year simulation period. \Cref{fig:favorablesliders} panel b shows the values of these same variables when $h=1$. $T_v=31.5, T_c=1.5,v_0=0.5, x_{R_0}=0.5,x_{P_0}=0.9,$ and the cost of expressing climate dissatisfaction, $d=1.5$.
    \end{minipage}
    \label{fig:favorablesliders}
\end{figure}

\subsection{Case 2: Increasing Homophily Improves Environmental Outcomes}
\begin{figure}[htb!]
    \centering
    \includesvg[width=1\linewidth]{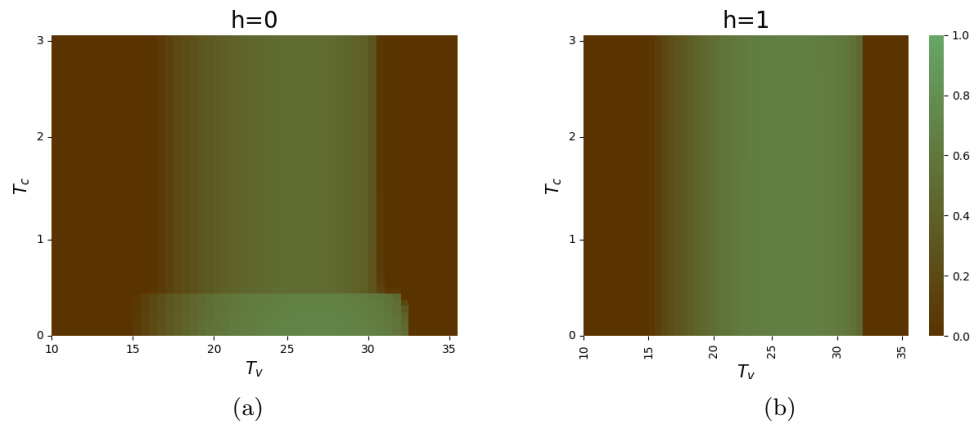}
    \caption{Final Forest Coverage Under Unfavorable Conditions For Mitigation}
    \begin{minipage}{15cm}
    \footnotesize
        Notes: The above figures depict heatmaps of forest coverage $v$ after 200 years under different ambient temperature $T_v$, critical temperature $T_c$, and homophily level $h$. \Cref{fig:unfavorableheatmap_full} panel a depicts the long term vegetative outcomes when $h=0$, while \cref{fig:unfavorableheatmap_full} panel b depicts the long-term outcomes when $h=1$.  To define a regime where mitigation is originally not favored, the perceived cost of climate change is low ($f_{\max}=4$), while the cost of expressing dissatisfaction over the state of the climate is high ($d=10$). $v_0=0.1, x_{R_0}=0.1, x_{P_0}=0.8$.
    \end{minipage}
    \label{fig:unfavorableheatmap_full}
\end{figure}

\begin{figure}[htb!]
    \centering
    \includesvg[width=1\linewidth]{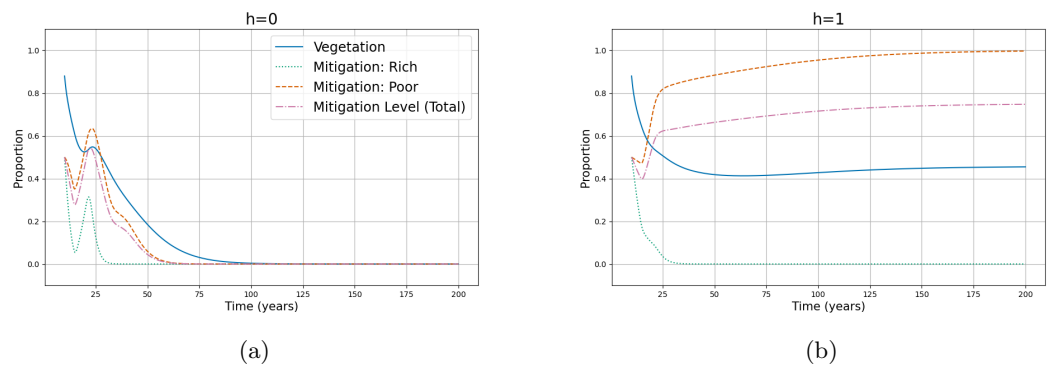}
    \caption{Trajectories Over Time Under Unfavorable Conditions For Mitigation}
    \begin{minipage}{15cm}
    \footnotesize
    Notes: The same initial vegetative state and mitigation proportions under different levels of homophily. \Cref{fig:unfavorablesliders_full} panel a shows the values of vegetation $v$, the mitigation proportions for the rich and poor groups, $x_R$ and $x_P$, respectively, as well as the total mitigation proportion, $x$, when $h=0$ over the course of the 200 year simulation period. \Cref{fig:unfavorablesliders_full} panel b shows the values of these same variables when $h=1$. $T_v=31.5, T_c=1.5, v_0=0.1,x_{R_0}=0.5,x_{P_0}=0.5$, and the cost of expressing climate dissatisfaction, $d=5$.
    \end{minipage}
    \label{fig:unfavorablesliders_full}
\end{figure}

Although the rich group is smaller in population size, its greater access to resources could give it greater influence on the climate through policies and technology, similar to real-world socioeconomic dynamics where the wealthiest nations have the power to largely shape international organizations such as the United Nations \cite{CFR2025} as well as acquire beneficial contracts for firms in the country \cite{BetterWorldCampaign2025}. \Cref{fig:unfavorableheatmap_full,fig:unfavorablesliders_full} depict patterns consistent with this disproportionate influence of the rich players. As seen in \cref{fig:unfavorableheatmap_full}, increasing homophily reduces the sensitivity of the final vegetation level to the critical temperature. Comparing \cref{fig:unfavorableheatmap_full} panels a and b, one can see a greater final vegetation proportion notably for critical temperature values greater than $1\degree$C. This implies that under low homophily, a society who is on average less sensitive to temperature rises will neglect climate action, leading to a worse environmental state. However, under high homophily, the group who has more to lose from a deteriorated climate are able to undertake mitigative actions on their own, leading a better climate outcome.

Additionally, \cref{fig:unfavorablesliders_full} paints a clearer picture of homophily stabilizing the vegetation levels. \Cref{fig:unfavorablesliders_full} panel a presents a simulation with low homophily. Here, lacking intergroup bias results in the poor group taking a stance of inactivity, alongside the rich group. However, in \cref{fig:unfavorablesliders_full} panel b, the high level of homophily results in poor group collusion to undertake mitigative actions, without the influence of the rich group. This subpopulation agreement in turn results in a vegetation proportion stabilizing well above 0, at about 0.45. Like was seen in the phase portraits, the tendency of the rich group to exert influence in low-homophily settings results in actions that more closely match only their incentives: accounting for greater homophily allows the poor group to collectively organize. 

One key finding to note is that negative climate outcomes can be avoided without complete participation in mitigative efforts. Although this has not been a focus of social-climate dynamic research, many climate activist groups have stressed the importance of smaller actions taken by non-state \cite{kuramochi2019global}, regional or sub-national groups \cite{weischermorganpatel2012}. This paper hence provides additional support to such forms of sectarian climate action.

\subsection{Alternative Parameter Sweeps}
To check the robustness of the results to many of the underlying parameters, this study performed additional simulations to gauge how the underlying social conditions, rather than simply temperature outcomes, impact the final vegetation proportion. 

\subsubsection{Clarifying the Relationship Between Homophily and Demographics}
\begin{figure}[htb!]
    \centering
    \includegraphics[width=0.5\linewidth]{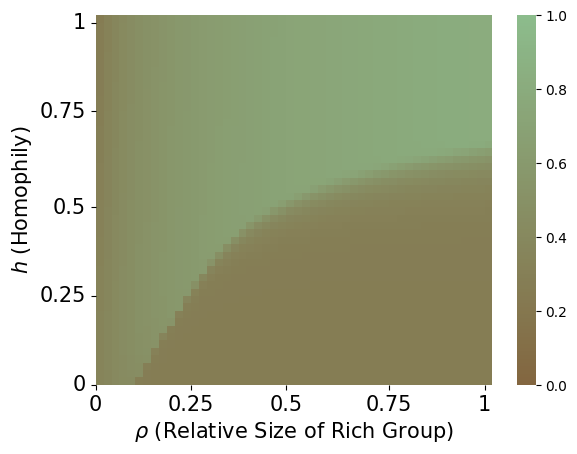}
    \caption{Final Forest Coverage Comparing Homophily and Group Size}
    \begin{minipage}{15cm}
        \footnotesize
        Notes: The above figure depicts a heatmap of forest coverage $v$ after 200 years under different levels of homophily $h$ and proportion of the population in the rich group $\rho$. $T_v=30$, $T_c=0.5$, $v_0=0.1$, $x_{R_0}=0.5$, $x_{P_0}=0.5$, $f_{\max}=4$, $d=10$.
        \end{minipage}
    \label{fig:homrhoplot}
\end{figure}

\begin{LineLabel}{R1-4}
Understanding the relationship between intergroup dynamics and group size is a critical marker of understanding group power and how opinions are disseminated. \Cref{fig:homrhoplot} depicts the relationship between the level of homophily and the size of the rich group relative to the entire population. When the rich group constitutes less than roughly 10 percent of the entire population, the final vegetative level does not depend on the level of homophily, equaling roughly 0.6 regardless of the homophily value. As the rich group increases in size, however, a homophily ``threshold'' appears, whereby a slight decrease in homophily places excessive weight on the opinions of the rich group, leading to a substantially worse climate outcome, differing between 0.3 and 0.7 in the given model specification. As the size of the rich group increases relative to that of the poor group, the level of homophily required to overcome this threshold also increases. The poor group (the group with more incentive to mitigate) must therefore be more insulated from the rich group (through greater homophily) to obtain the same long-term climate outcomes as the size of the rich group increases.
\end{LineLabel}

\subsubsection{The Impact of More Effective Mitigation}

\Cref{fig:salpha_full} investigates the relationship between costs of mitigation and effectiveness of mitigative actions. 
\begin{LineLabel}{s}
    Due to knowledge or technological improvements, mitigation may become a task which requires less effort in the future. These knowledge or technological improvements likely will require significant funding and research. Hence, our model assumes that it is the rich group who has the ability to generate these improvements, while the poor group is unable to do so.
\end{LineLabel}
\begin{LineLabel}{R1-8.1}
Firstly, increasing homophily increases the range of initial cost ratios that result in maximized vegetative outcomes. As homophily increases, groups are less motivated to compare their costs of mitigation, as they are less considerate of the conditions of the other group. As $s$ increases, the threshold between high and low vegetation levels occurs at lower ratios of mitigation costs.
\end{LineLabel}
~Thus, as the relative effectiveness of mitigative 
\begin{LineLabel}{R2-vi}
   action 
\end{LineLabel}
~(the mitigative effort from the rich group) increases, the rich group only chooses to mitigate (leading to the light green area) when they are given a lower relative cost of mitigation. 
The sentiment in favor of each country ``doing their own share'' has been seen gaining traction in current discourse. Hence, this further highlights the divide in incentives between rich and poor groups when it comes to mitigative activity.

\begin{LineLabel}{R1-8.2}
    Finally, the effectiveness of rich mitigative actions appears to itself have a threshold for additional usefulness. Regardless of the homophily value, the downward slope of the vegetation threshold flattens out around an $s$-value of 6. Hence, for highly effective rich mitigative actions, the relationship between the ratio of mitigation costs and the vegetation threshold (that the threshold occurs at lower cost ratios as $s$ increases) is wiped out. The power of each group ``doing their own share'' is weakened in favor of collective action as the rich group has significant capabilities to mitigate.
\end{LineLabel}

\begin{figure}[htb!]
    \centering
    \includesvg[width=1\linewidth]{Figures/s_sweepcombined.svg}
    \caption{Final Forest Coverage Allowing For More Effective Mitigative Action}
    \begin{minipage}{15cm}
        \footnotesize
        Notes: The above figures depict heatmaps of forest coverage $v$ after 200 years under different relative mitigation costs for the rich group ($\alpha_{R_0}/\alpha_{P_0}$), effectiveness of mitigation from the rich group $s$, and homophily levels $h$. \Cref{fig:salpha_full} panel a depicts the long term vegetative outcomes when $h=0$, while \cref{fig:salpha_full} panel b depicts the long-term outcomes when $h=1$. An increase in $s$ can be understood as improvements in mitigative technology which belong exclusively to the rich group, but provide benefit to all individuals. $T_v=30, T_c=0.5, v_0=0.1, x_{R_0}=0.5, x_{P_0}=0.5$.
    \end{minipage}
    \label{fig:salpha_full}
\end{figure}

\subsubsection{The Cost and Immediacy of Warming Effects}
To analyze how people responded to their perceived effects of climate change (not merely the observed state of the forest), this study compared the maximum warming cost to the nonlinearity of warming costs. \Cref{fig:fmaxw_full} clearly shows much greater robustness of the system to greater warming costs or nonlinearity effects in regimes with greater homophily. 

\begin{figure}[htb!]
    \centering
    \includesvg[width=1\linewidth]{Figures/hetfmaxsweepcombined.svg}
    \caption{Final Forest Coverage Under Changing Cost and Immediacy of Warming Effects}
    \begin{minipage}{15cm}
    \footnotesize
    Notes: The above figures depict heatmaps of forest coverage $v$ after 200 years under different nonlinearity of warming costs $w$, maximum warming costs $f_{\max}$, and homophily levels $h$. \Cref{fig:fmaxw_full} panel a depicts the long term vegetative outcomes when $h=0$, while \cref{fig:fmaxw_full} panel b depicts the long-term outcomes when $h=1$.  $T_v=31.5, T_c=1.5, v_0=0.1, x_{R_0}=0.5,x_{P_0}=0.5$.
        
    \end{minipage}
    \label{fig:fmaxw_full}
\end{figure}
In the no homophily regime (\cref{fig:fmaxw_full} panel a), players must experience a very large warming cost ($f_{\max}>6$) - and have that cost be incurred rather slowly ($w<3$) - to be incentivized enough to mitigate and avoid climate catastrophe. However, in the high homophily regime (\cref{fig:fmaxw_full} panel b), these barriers are almost completely removed, allowing people to act while perceiving ``weaker'' climate impacts much more quickly.

Our model exhibits a potentially counterintuitive effect. While it might be natural to encourage immediate action minimizing warming, \cref{fig:fmaxw_full} suggests that such activism (increasing the perceived nonlinearity of warming costs) actually reduces the range of scenarios where a positive climate outcome is achieved. To analyze this further, we must consider our warming function. When $-7.5v(t)+7.5v(t-10) < T_c$, ${-w[-7.5v(t)+7.5v(t-10)-T_c]}>0$. In this scenario, increasing $w$ decreases $f(T_f)$ in \cref{eq:ftf}. For this relation to hold, $v(t)-v(t-10)>-\frac{2T_c}{15}$. For the results shown in \cref{fig:fmaxw_full}, this means that $v(t)-v(t-10)>-0.2$, or that the vegetation does not decay by more than $0.2$ over 10 years. While the sigmoid function is allowed to ``flip'' throughout our simulations, these results suggest that for most of the years of the simulation, the inequality $v(t)-v(t-10)>-0.2$ holds.

\subsubsection{Costs of Mitigation and Social Norms}

\begin{figure}[htb!]
    \centering
    \includesvg[width=1\linewidth]{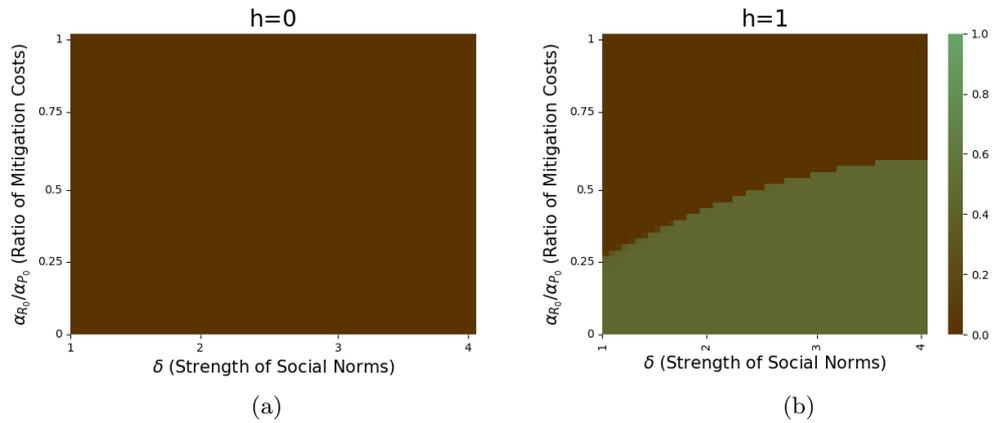}
    \caption{Final Forest Coverage Under Changing Cost of Mitigation and Social Norms}
    \begin{minipage}{15cm}
    \footnotesize
    Notes: The above figures depict heatmaps of forest coverage $v$ after 200 years under different relative mitigation costs for the rich group ($\alpha_{R_0}/\alpha_{P_0}$), social norm strength $\delta$, and homophily level $h$. \Cref{fig:alphadelta_full} panel a depicts the long term vegetative outcomes when $h=0$, while \cref{fig:alphadelta_full} panel b depicts the long term outcomes when $h=1$. $T_v=31.5, T_c=1.5, v_0=0.1, x_{R_0}=0.5,x_{P_0}=0.5$.
        
    \end{minipage}
    \label{fig:alphadelta_full}
\end{figure}

When discussing strategies to combat climate change, two of the most frequently mentioned pathways involve investments in green energy and working to change societal opinions on a warming planet. Hence, we examine these factors simultaneously, comparing regimes under different levels of homophily. The results are shown in \cref{fig:alphadelta_full}. 

Unsurprisingly, the results of these simulations are highly sensitive to homophily: homophily is strongly correlated to the relevance of social norms. 
\begin{LineLabel}{R1-8.3}
Note a threshold existing in \cref{fig:alphadelta_full} panel b, whereby if the ratio of mitigation costs exceeds the threshold, barren vegetation results, but if the ratio does not exceed the threshold, a vegetation level of roughly 0.45 is able to proliferate. 

In \cref{fig:alphadelta_full} panel b, also note that as the strength of social norms increases (making it more favorable for individuals to join with the consensus of their group), the threshold of the ratio of mitigation costs between rich and poor groups increases. This implies that as the benefit to joining with the group consensus increases, the importance of relative costs for the group decreases, leading to a proliferated vegetative outcome in a greater number of cost regimes.
\end{LineLabel}
This is an encouraging but not unsurprising result, as individuals feel a greater sense of social responsibility, they are also willing to initiate more monetary support for climate-friendly goals.

\section{Conclusion}

Collective action is critical to minimizing climate damages~\cite{levin2022governance}. While natural inclination leads us to assume that promoting minimum homophily is the way to get us there, this study shows that this is not always the case. When a group has little to no incentive to mitigate (as is the case with the rich group, who bears little of the costs of climate inaction), dialogue may be futile. Coupling a forest dieback model \cite{fushu:23} with more robust population dynamics \cite{menard2021}, we see that the rich group's inaction drags down the poor group's mitigation proportion when communication is open between the two groups ($h=0$). However, because the poor group bears more of the costs of mitigation, it is commonly in their best interests to mitigate. In this case, creating an ``echo chamber'' through high homophily ($h=1$) promotes collective climate action among the poor group, which results in more vegetative states.

\begin{LineLabel}{R1-1}
 This paper confirms that low homophily creates opinion convergence, which can be driven either towards climate action or inaction, typically by the sentiment of the richer/more powerful group. High homophily, on the other hand, leads to opinion divergence, which results in better climate outcomes than consensus inaction, but worse outcomes than consensus action. Such non-consensus outcomes have been of little interest previously, but are beginning to gain traction as global action on climate change has appeared very difficult to achieve \cite{RobiouDuPontEtAl2025}.
\end{LineLabel}

Additionally, improving a sense of social responsibility for all players unsurprisingly increases the range of climate scenarios preserving some level of vegetation. Furthermore, technological improvements actually increase the range of barren climate scenarios, casting doubt on technological improvements being able on their own to lead to lasting warming mitigation (because they fail to address underlying motivations to mitigate).

Comparing regimes, it is clear that there are thresholds where crossing over results in movement between a green and a barren regime. Analyzing the parameter models in this theoretical system does not necessarily lend itself to direct physical interpretations, but \cite{ritchie2021overshooting} and others do find evidence for tipping points of such systems. This model, while not directly investigating tipping points, nonetheless clearly finds evidence for them, as there are drastic differences in vegetative outcomes after the 200 year simulation period due to small changes in the underlying parameters.

While this model currently assumes that players accurately understand climate dangers and impacts, a further investigation of individual perception of risk~\cite{wang2024evolution}, and how that changes in relation to a changing climate, would make this model more robust to changing sentiments as well as a changing climate. Additionally, further analysis of tipping points and overshooting climate targets is another related detail which should be investigated further in this context.

In sum, we study a social-climate system model that incorporates into the die-back forest model with groups of individuals whose mitigation behavior is influenced by social factors including wealth inequality and homophily. We determine dynamic regimes that lead to either successful mitigation or failure to prevent climate disasters. Our proof-of-concept model sheds light on understanding how the intrinsic nonhomogeneity of human societies, from `the poor vs the rich'~\cite{wang2010effects} to `us vs them'~\cite{fu2012evolution}, shapes behavior and attitudes in collective-risk dilemmas beyond climate change~\cite{wang2009emergence}.   

\ack{We are indebted to Mark Lovett and members of the Fu Lab for their incredible guidance, as well as visitors to the Wetterhahn and Dartmouth Mathematics Department symposiums for their helpful comments.}

\funding{L.W. is incredibly grateful to the Mellam Family Foundation for funding of this research.}

 \roles{L.W.: Conceptualization, Formal analysis, Investigation, Methodology, Software,Visualization, Writing – original draft; T.Z.: Conceptualization,  Investigation, Methodology, Project administration, Supervision, Writing – review \& editing; F.F.: Conceptualization, Supervision, Writing – review \& editing.}


\data{The code used to create the simulations in this paper can be found at the following link: \url{https://github.com/dartmouth/SocialClimateDynamics}.
}

\suppdata{This article has no additional data. All relevant data has been included in the main text.}

\bibliographystyle{unsrt}
\bibliography{main}

\end{document}